\documentclass[twocolumn]{aastex62}

\newcommand       \mum           {\,{\rm \mu m}}

\graphicspath{{./}{figures/}}

\shorttitle{}
\shortauthors{}

\begin{document}

\title{Discovery of two nearby post-T Tauri stellar associations}
\correspondingauthor{Chao Liu}
\email{liuchao@nao.cas.cn; mfang@email.arizona.edu; jmliu@nao.cas.cn}

\author{Jiaming Liu}\footnote[1]{Jiaming Liu, LAMOST Fellow.}
\affiliation{Key Lab of Optical Astronomy, National Astronomical Observatories, CAS, Beijing 100101, China}

\author{Min Fang}
\affiliation{Department of Astronomy, University of Arizona, 933 North Cherry Avenue, Tucson, AZ 85721, USA}

\author{Chao Liu}
\affiliation{Key Lab of Space Astronomy and Technology, National Astronomical Observatories, CAS, Beijing, 100101, China}
\affiliation{University of Chinese Academy of Sciences, Beijing 100049, China}




\begin{abstract}
In this work we report the discovery of 2 new stellar associations in close vicinity of the Sun at roughly 180 and 150 pc. These two associations, named as u Tau assoc and e Tau assoc,  were detected based on   their clustering in a  multi-dimensional parameter space including $\alpha, \delta, \mu_{\alpha}, \mu_{\delta}$ and  $\varpi$ of {\it Gaia}. The fitting of pre-main-sequence model isochrones in their color-magnitude diagrams suggests that the two associations are of about 50 Myr old and the group members lower than $\sim$0.8~$M_{\odot}$ are  at the stage of post-T Tauri.


\end{abstract}

\keywords{Young star clusters, Pre-main sequence, Pre-main sequence stars}



\section{Introduction} \label{sec:introduction}


 Stellar associations, as fundamental blocks of our Milky Way, will help us to understand the formation and evolution of the structures of the Milky Way. Nearby young associations are particularly important, and are excellent laboratories for studying
 the initial mass functions (IMF) in extremely low mass range \citep{Gag2018}, formation and early evolution of planetary systems and brown dwarfs (BD) \citep{Chauvin2015}, since young objects of sub-stellar and planetary mass range are comparatively bright and easy to be detected.  However, due to the low stellar density and large extension in the sky, nearby young associations and their members are hard to be identified. Members of them are newly formed in same molecular cloud and haven't been significantly perturbed by the Galaxy, thus they usually share some common properties such as age, chemical composition, distance and kinematics \citep{Son2003,Tor2008,Gag2018}. Therefore, these stars will usually show significant concentration in multi-dimensional astrometric space and be identified as over-densities from its surrounding field stars \citep{Fur2019}.

 Recently, \emph{Gaia} DR2 data, which provides position and G band (330-1050 nm) photometric data down to magnitude 21 for 1.7 billion stars, including parallax and proper motions for over 1.3 billion stars and radial velocity for stars brighter than 13 magnitude in G band \citep{GAIA2016,GAIA2017,GAIA2018}, will definitely promote the membership completeness of current stellar associations and boost the discovery of new stellar associations.
Here, we report the discovery of two new young associations near Taurus. 

This work is arranged as follows. In $\S$2, we describe the data and refine the membership of the associations. The population properties will be discussed in $\S$3. Followed by a brief discussion in $\S$4, and finally summary and conclusion in $\S$5.

\section{data and membership}
\subsection{Data Selection}

\begin{figure*}
\centering
\includegraphics[height=10cm]{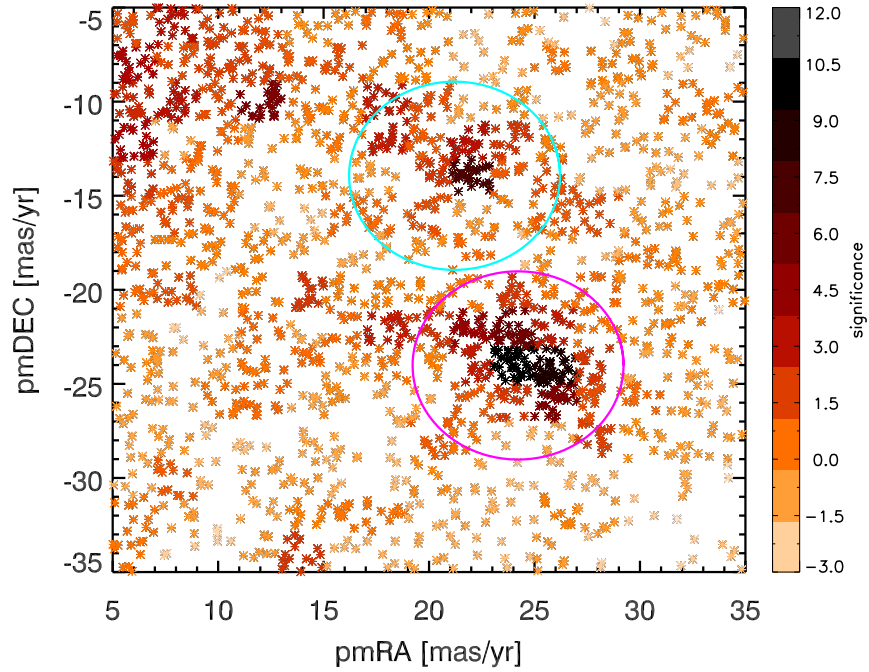}
\caption{The location of the two associations in $\mu_{\alpha}$, $\mu_{\delta}$ space. Based on their locations in the proper motion, stars are divided into a series bins of 2 mas/yr$\times$2 mas/yr, while colors of symbols in each bin denote the significance of them. Cyan and purple circles are 5 mas/yr region around the center of the potential associations.}
\label{cube}
\end{figure*}

During a study about the young stellar associations of Taurus (Liu J. et al. 2020, in preparing) with {\em Gaia} DR2 data, we notice that
apart from those known associations, there are two likely stellar associations near Taurus, located at roughly 150 and 180 pc that haven't been realized before. The two candidates of associations that roughly centered to ($+$21.21,$-$13.94) and ($+$24.22,$-$24.02) of ($\mu_{\alpha}, \mu_{\delta}$), are clearly notable from their surrounding field stars as outstanding over-densities in the proper motion space (see figure \ref{cube}). Stars employed in this plot are selected by 50$^{\circ}$ to 65$^{\circ}$ of $\alpha$, 0$^{\circ}$ to 20$^{\circ}$ of $\delta$, 15 to 35 mas/yr of $\mu_{\alpha}$, -30 to -5 mas/yr of $\mu_{\delta}$ and 100 to 200 pc of distance. As the two associations are tightly clustered in the proper motion space, to refine their memberships, the data we selected are based on the following criteria: (1) all those sources around the center of these two associations with a radius of 5 mas/yr from the proper motion space, slightly larger than the radii of them; (2) a parallax quality of $\sigma_{\varpi}/\varpi \le 0.1$; (3) the flux error of $G_{RP}$ smaller than 5\%; considering that the typical distance extension for nearby young associations are usually $\sim$20~pc \citep{Gag2018}, (4) the distance cut is set to be 100-200 pc from the Sun. In total we have 448 stars.

\subsection{Membership}

\begin{figure*}
\centering
\includegraphics[height=10cm]{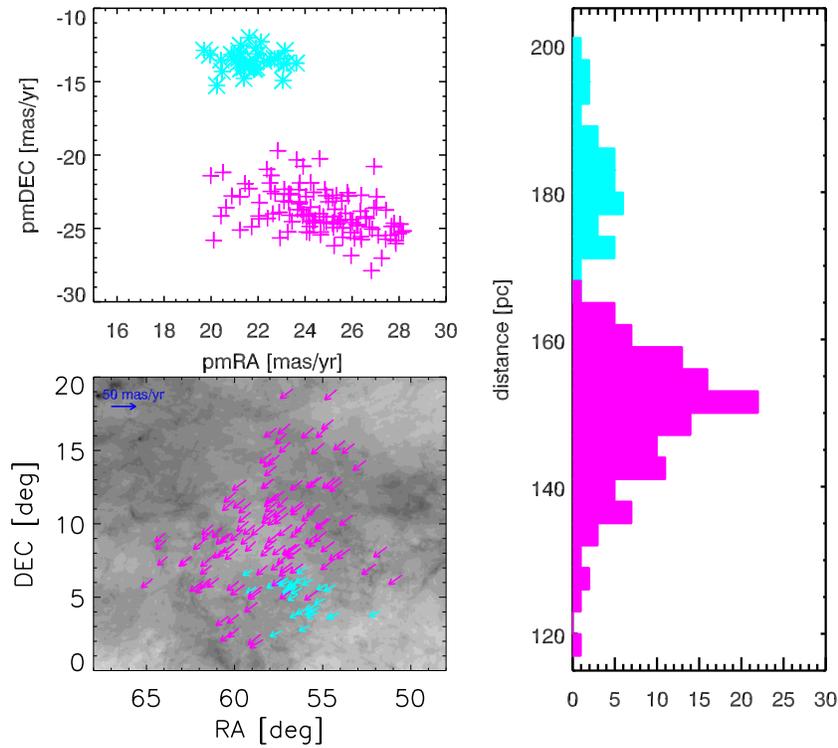}
\caption{The location of the two associations at $\alpha$, $\delta$ space (lower-left panel), $\mu_{\alpha}$, $\mu_{\delta}$ space (upper-left panel) and distance distribution (right panel). Background contour in the lower-left panel are the 350$\mum$ dust emission map of Planck.}
\label{FOF}
\end{figure*}

Membership refinement by their concentration in multiple astrometric space 
is the most commonly used method for associations showing clear over densities in the astrometric space. 
Considering members of the nearby associations 
are usually loosely concentrated in the spatial space, 
a ``Friend-of-Friend'' (``FOF'') algorithm of ``ROCKSTAR''
\footnote{ROCKSTAR is based on adaptive hierarchical refinement of 
friends-of-friends (FOF) groups in six phase-space dimensions,
which allows for the tracking of high number density clusters, 
and divided all the stars into several groups excluding those stars that
could not be vested in the star aggregates.
It is designed to find out outlier structures that tightly connected in the 6 dimension
space, as {\it Gaia} DR2 only provides radial velocity for bright stars with G band magnitude brighter than 13.0.
Thus, to adopt ROCKSTAR, we set radial velocity as zero for all sample stars and keep the 
other 5D coordinates as $\alpha, \delta, \mu_{\alpha}, \mu_{\delta}$ 
and distances given by \cite{Bai2018}.} 
is adopted \citep{Beh2013}.
Based on the sample stars input, ROCKSTAR will automatically modify
the linking-space between members of ``friend'' stars, 
divide them into several groups, and label those isolated individual stars out. 
Thus, with these 448 stars we selected above, we ruled out the
surrounding field stars by eliminating isolated stars at each run time.
After 3 iterations of the ``ROCKSTAR'', 
The left stars are spatially concentrated in two groups (see figure \ref{FOF}),
which contain 35 and 119 members, respectively. 
We regard them as u Tau assoc and e Tau assoc (see table \ref{LFL1} 
and \ref{LFL2} for detail information) hereafter. A reliability test of ROCKSTAR
in refining memberships of the 2 associations is proven in the appendix. The 
ROCKSTAR can find out $\sim 95\%$ of the group members at a purity level of $\sim 90\%$, 
proved that the ROCKSTAR method is a effective way in refining memberships of Tau-assocs kind.
Figure \ref{FOF} shows the locations of the associations,
the cyan asterisks indicate the stars of u Tau assoc 
and the purple crosses represent the members of
e Tau assoc. The association u Tau assoc, located at roughly 180 pc from the Sun, 
is tightly concentrated in both $\alpha$ and $\delta$ 
(centered at 56.0$^{\circ}$, 5.46$^{\circ}$) and proper motion space.
The e Tau assoc is located at 150 pc from the Sun, tightly concentrated in the proper motion space but
largely extended in the $\alpha$ and $\delta$  space (centered at 57.69$^{\circ}$, 10.16$^{\circ}$).

\subsection{Spectra Data}

The spectra data adapted in this work is taken from LAMOST. The Large Sky Area Multi-Object Fibre Spectroscopic Telescope(LAMOST), is a 4m Schmidt telescope of the National Astronomical Observatories, located at Xinglong Observing Station, China. With 4000 fibers on board the focus, the LAMOST can observe nearly 4000 optical spectra in a 20 square degrees a time \citep{Cui2012}. Since 2012, the LAMOST has released its surveying data 7 times. The data that engaged in this work are taken from the recently released data DR5. LAMOST DR5 provides over 9 million spectra as well as a catalog of general stellar parameters derived from the spectra for over 5 million stars. With LAMOST data, we will discuss some properties of the associations later.

\section{population properties}
\subsection{Convergent Point}

As a projection effect, the co-moving members of an association usually 
converge to a certain point of the celestial sphere, known as convergent
point \citep[CP;][]{Gal2012}. The CP point, ever since its first presented by 
Bohlin in 1916 \citep{Sma1968}, has also been used as an important tool for testing
membership of associations. Since then a lot of different methods 
have been introduced for its solution \citep{Jon1971,de1999,Gal2012}.
In order to confirm that these 2 new associations are not from the same group, we
derive their convergent points of the equatorial coordinates.
The CPs for u Tau assoc and e Tau assoc are (98.06$^{\circ}$, 
-19.32$^{\circ}$)$\pm(0.67^{\circ},0.36^{\circ})$ 
and (108.39$^{\circ}$,-33.23$^{\circ}$)$\pm(0.54^{\circ},0.28^{\circ})$, respectively. 
The difference of the convergent points is evident that
the two associations, although close in the location and share similar properties, are indeed two separated ones.

\subsection{Age and Mass}\label{agemass}

\begin{figure*}
\centering
\includegraphics[height=12cm]{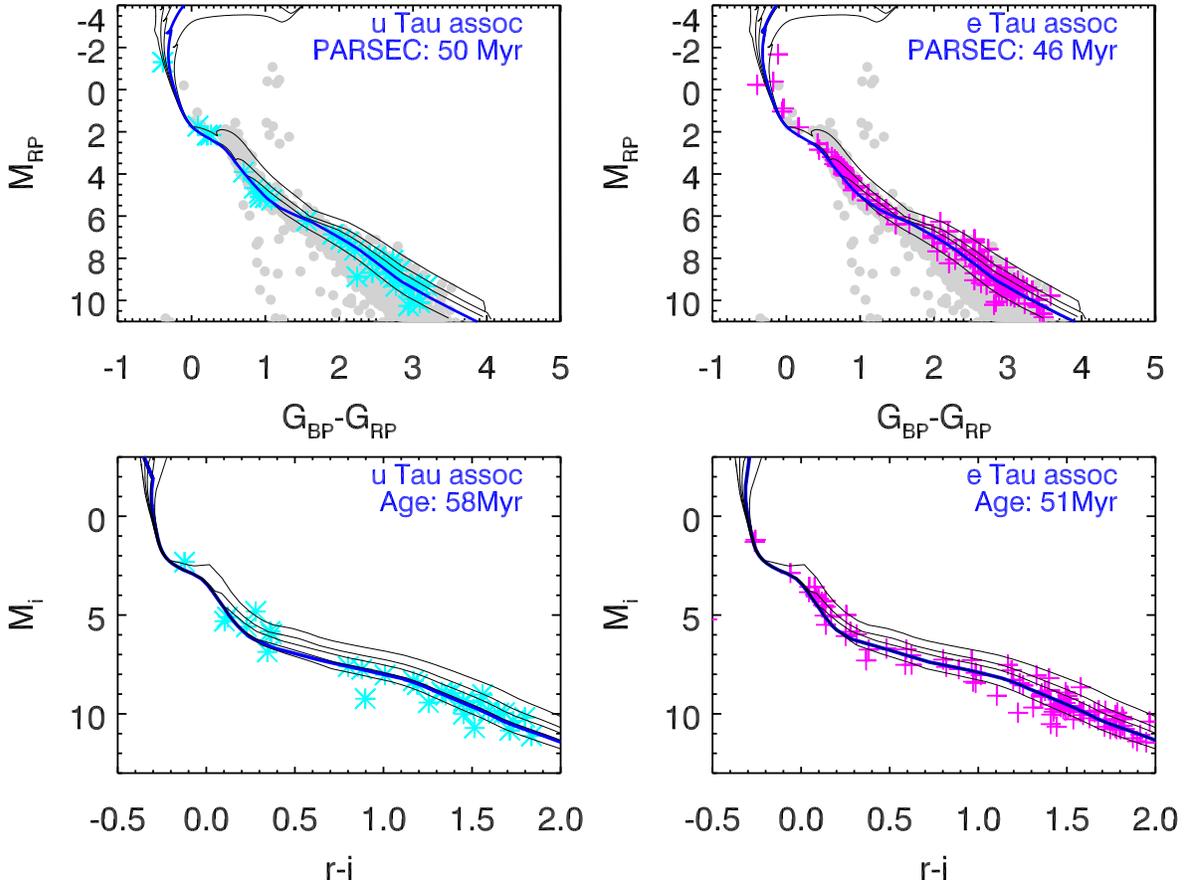}
\caption{The color-magnitude diagram of Tau-assocs. 
Gray dots are the foreground main-sequence stars in the same direction. Cyan asterisks denote the u Tau assoc, while purple crosses indicate members of e Tau assoc. The blue solid line are the best fitted isochrones of PARSEC (upper panels) and model of \cite{Bel2014} (lower panels). The thin black lines from the top to the bottom in each panel denote isochrones of 10, 20, 30, 50 and 100 Myr.
}
\label{CMD}
\end{figure*}

Figure \ref{CMD} shows the CMDs for the 2 associations. The gray dots are the 
foreground stars in the same direction of them but within 100 pc
from the Sun, and are regarded as main-sequence stars here. 
The foreground stars are selected based on the following criteria:
1. stars in the region of 0-20$^{\circ}$ of $\alpha$ and 50-65$^{\circ}$ of $\delta$;
2. $\sigma_{\varpi}/\varpi \le 0.1$;
3. the flux error of $G_{RP}$ smaller than 5\%;
4. stars in 100 pc from the Sun, in total we have 1934 stars. Asterisks and crosses represent for the members of the associations u Tau assoc and e Tau assoc, respectively.
Interstellar extinction is corrected using the Galactic average extinction law of $R_{V}=3.1$, where $R_{V}=A_{V}/E(B-V)$ is the total to selective coefficient. Since the $V$ band interstellar extinction is about 0.7-1.0 mag/kpc in solar neighbourhood \citep{Got1969,Mil1980,Wan2017}, a mean value of 0.85 mag/kpc is taken here. Then for each individual star at distance D, its V-band extinction $A_{V}$ should be $0.85\times D$. Finally, with the extinction coefficient for {\it Gaia} $G_{BP}$ and $G_{RP}$ bands of $R_{V}=3.1$ that provided by \cite{Wan2019}, the $G_{BP}$ and $G_{RP}$ extinction for each individual star of both foreground main-sequence stars and the associations members are corrected. 
The mean $V$ band extinctions for u Tau assoc and e Tau assoc are 0.154 mag and 0.127 mag, respectively.

The blue dashed line in each panel are the best fitted isochrone of PARSEC \citep{Bre2012} of solar metallicity. The fit is done as follow: for each isochrone of certain age, select those stars within the color and absolute magnitude range of the isochrone. Distances of an isochrone to each of these stars are derived, as the number of stars inside the color and magnitude range is different from isochrone to isochrone, thus the isochrone with smallest mean distance to stars are regarded as the best-fit.
The result showed that these two associations are both young, 50 Myr for u Tau assoc and 46 Myr for e Tau assoc. Actually, as a result of core fusion and convection, surface Lithium abundance will deplete with time, and can also be used to estimate age of PMS stars \citep{Her1965,Wey1965}. But this method is not appropriate for LAMOST low-resolution spectra, whose resolution is $\sim 1800$, hard to deduce the Lithium abundance accurately. Therefore Li-based method is not considered in this work. 

In order to better confirm the result of the age estimation, we alternatively considered the semi-empirical pre-main-sequence model isochrones of \cite{Bel2014} for SDSS bands, which they derived base on the Pisa PMS isochrones of \cite{Ton2012}.
Realizing that over half of the stars lack of SDSS observations, thus we cross-match the members of Tau-assocs to PanSTARRS \citep{Cha2016}, of which only 2 stars of e Tau assoc have no counterparts. But to control the security of the model fit, only those stars with good-quality measurements of PanSTARRS (quality flag of 4, 8 and 16) are conserved (32 stars of u Tau assoc and 106 stars of e Tau assoc). Then we transferred these photometric data of PanSTARRS to SDSS bands with method of \cite{Tonr2012}, of which the transition accuracy is better than 0.002 and 0.003 mag in $r$ and $i$ band, respectively. The model fit result is also shown in figure \ref{CMD}, and the age are estimated to be 58 and 54 Myr for u Tau assoc and e Tau assoc respectively, well agree with the age estimation of PARSEC.

\begin{figure}
\centering
\includegraphics[height=8cm]{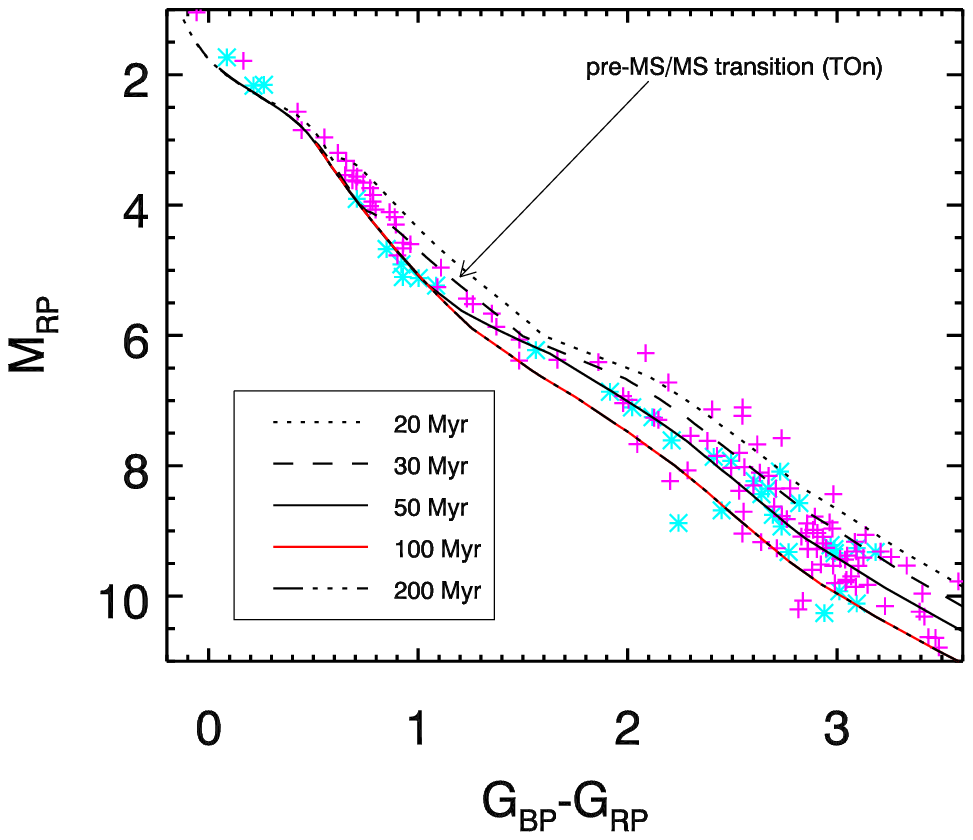}
\caption{Pre-main sequence tracks of different ages given by PARSEC, also plotted are the members of Tau-assocs with symbols defined the same as before. }
\label{TOn}
\end{figure}

Besides, the transition point (e.g.\,the Turn-On point: TOn) from PMS to main-sequence (MS) can also be used to estimate age of young stellar groups. 
Because the TOn point on the CMD will become flatter and drop to the main-sequence tracks as the stellar group growing old \citep{Cig2010}. Therefore, TOn can also serve as an indicator of age for clusters that sufficiently young to contain PMS members.
In light of this, the TOn of PARSEC model is also considered as a validation of our age estimation. The result of this approach is shown in figure \ref{TOn}. TOns of the Tau-assocs are both consistent with that of the 50 Myr, with the TOn point roughly correspond to a $\sim 0.8 M_{\odot}$ star.

\begin{figure}
\centering
\includegraphics[height=6cm]{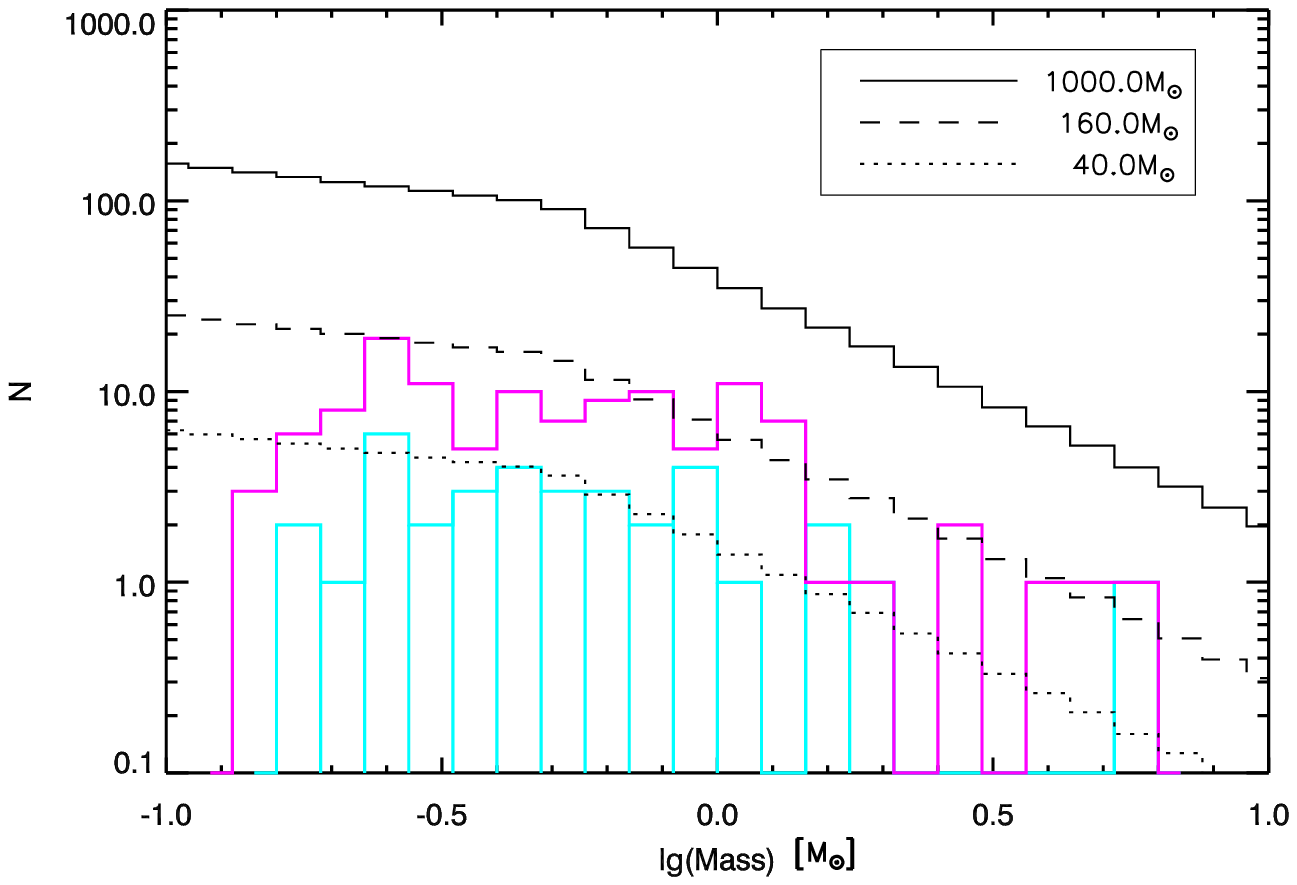}
\caption{Mass estimation of the associations based on the IMF model of \cite{Kro2001}.
}
\label{IMF}
\end{figure}

With the best fitted isochrones of PARSEC, stellar masses are then estimated. The mass range of the associations demonstrate that most of their members are low mass (sub-solar mass) with a few of them has masses larger than solar mass, and an upper limit of $\sim 6.0 M_{\odot}$.
Based on the masses of stars, we fit the present-day mass functions to a series of IMF (initial mass function) of \cite{Kro2001}, and the total mass of the 2 associations are estimated to be $\sim40$ and $\sim$160 $ M_{\odot}$ respectively for u Tau assoc and e Tau assoc (see figure \ref{IMF}).

\subsection{Radial Velocity and Metallicity}

\begin{figure}
\centering
\includegraphics[height=6cm]{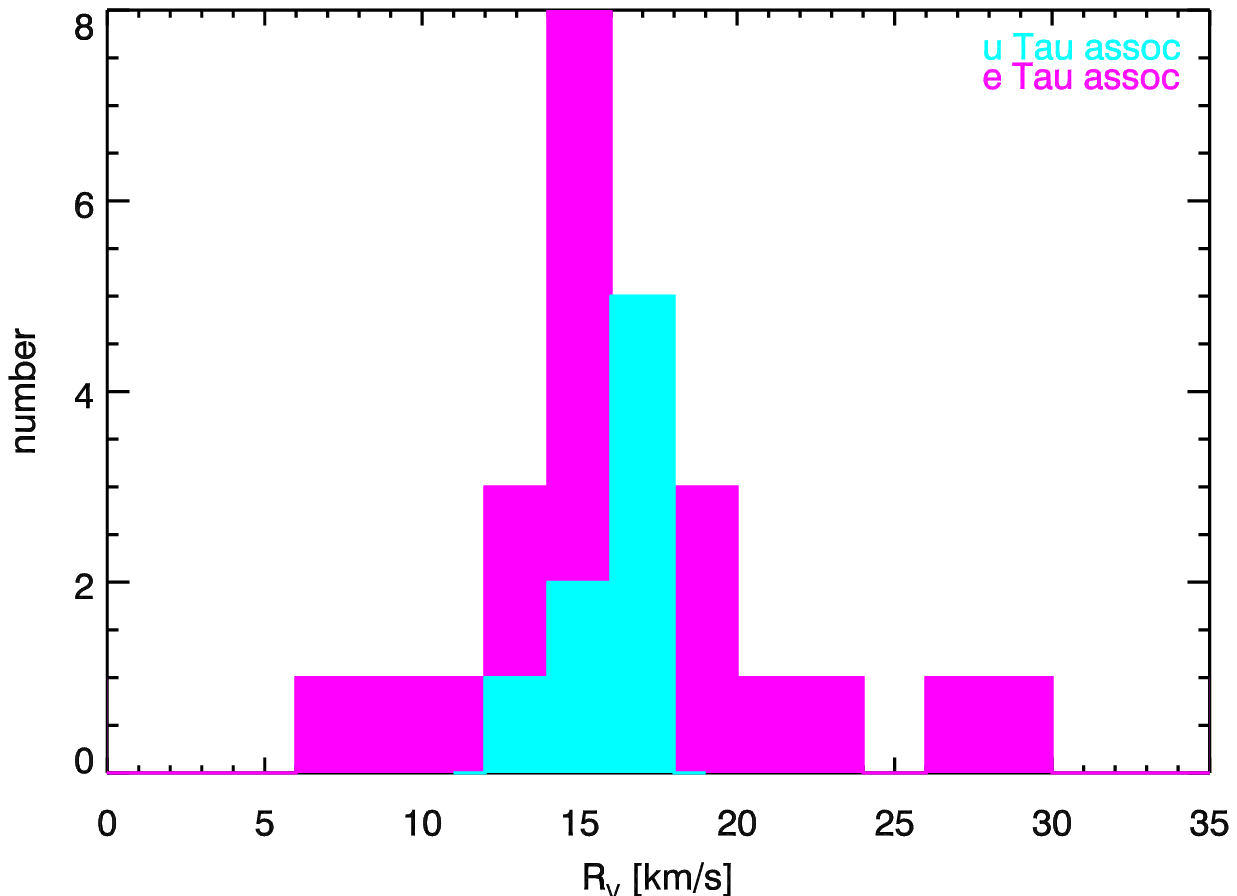}
\caption{Radial velocities histogram of Tau-assocs. Cyan for u Tau assoc, and purple for e Tau assoc.}
\label{Rv}
\end{figure}

Radial velocity is an important parameter of kinematics, especially for stars of the associations. Since {\em Gaia} DR2 only provided the radial velocities for bright stars, 7 star in u Tau assoc and 29 in e Tau assoc have {\em Gaia} released radial velocities. We also searched members of the associations in the LAMOST DR5, and 23 stars of them are found in the LAMOST DR5 catalogue, but only 1 of them is new. In a word, we have 8 members of u Tau assoc and 29 members of e Tau assoc with radial velocity information. The histograms of the radial velocity for the associations are plotted in figure \ref{Rv}. A rough examination showed that the mean value of the radial velocity of u Tau assoc is 16 km/s with dispersion of 2 km/s, while for e Tau assoc the average value is 15 km/s and a dispersion of 5 km/s. 

Besides radial velocity, LAMOST DR5 also provided metallicities for 7 stars in u Tau assoc and 16 in e Tau assoc. The mean metallicity for u Tau assoc and e Tau assoc are 0.03 and 0.03 dex, respectively.

\section{discussion}\label{discuss}

Associations, especially those nearby associations are the ideal laboratory for studying stellar kinematics and evolution of stars in groups, therefore, efforts have been taken for searching more new associations, especially after the data release of {\em Gaia}.  Quite soon after the first data release of {\em Gaia}, \cite{Oh2017} have applied a marginalized likelihood ratio test to the Tycho-{\em Gaia} Astrometric Solution (TGAS), searching for co-moving pairs from field stars. From 10,606 unique stars, by a connection radius of 10 pc, they find 13,085 co-moving star pairs. Later, \cite{Fah2018} pushed the work forward and reexamined the result of \cite{Oh2017} with the bayesian method BANYAN $\Sigma$, apart from those associations of already known, they also reported over 20 potential new stellar groups. Among their potential stars in groups, 10 members of the association e Tau assoc of this work are included, but were separated in 3 different groups. A likely reason for this is that for the very time of their work, TGAS is short in stars, as e Tau assoc is largely extended in the space, it's natural that the members were not linked together and divided into several parts. On the other hand, their work also confirmed the fact that these stars are clustered in groups.





A 2 arcsec radius cross-match with SIMBAD show that these 2 associations in total contains 43 counterparts, including 3 B3, 4 B9, 1 A1, 3 A5, 8 F type, 7 G type, 3 K type, 1 M type stars (13 with no spectra type information). This is consistent with the mass estimation results in section \ref{agemass} that the associations of this work contains a few stars of several solar masses. 
\cite{Tet2011} published a catalogue of young {\it Hipparcos} stars within 3 kpc from the Sun. 4 of the 43 SIMBAD sources in this work are included, and the age are estimated as 5.7, 7.0, 13.4 and 37.8 Myr, which are younger than the ages of this work. The difference is understandable since both works using different distances. In \cite{Tet2011}, they adopted the distances from {\it Hipparcos}, with seven sets of evolutionary models, they derived the median ages from them (see details in \cite{Tet2011}). For two sources with ages of 5.7 or 7.0~Myr in \cite{Tet2011}, there are large difference between the distances from {\it Hipparcos} and {\it Gaia} with the distances from {\it Hipparcos} being 26\% and 16\%, respectively, larger than those from {\it Gaia}. This can result in higher luminosity and younger ages in \cite{Tet2011} than this work. Furthermore, the age of each association in this work is estimated as a whole from CMD or the TOn, which should provide a better age estimation than just using individual stars.


Early in 1978, Herbig introduced post T Tauri stars (PTTS) in order to explain the lack of ``older'' pre-main sequence T Tauri stars with age older than $5-10$ Myr in star forming regions \citep{Her1978}. As the distinct characteristics of T Tauri stars, e.g. strong H$_{\alpha}$ emission, high surface lithium abundance, irregular variability, and color excess of infrared as well as X-ray emission \citep{Wal1986} are absented at stage of zero-age main sequence (ZAMS),  he argued that the PTTS should show intermediate or moderate value of these characteristics. \cite{Jen2001} thought that PTTS are stars older than T Tauri but haven't reach zero-age main sequence (ZAMS), and discussed the properties of them in his work. He stated that to certificate the validity of PTTS, apart from age estimation, the validity of PTTS should be further confirmed by at least one or two characteristics of T Tauri stars.

\begin{figure}
\centering
\includegraphics[height=7cm]{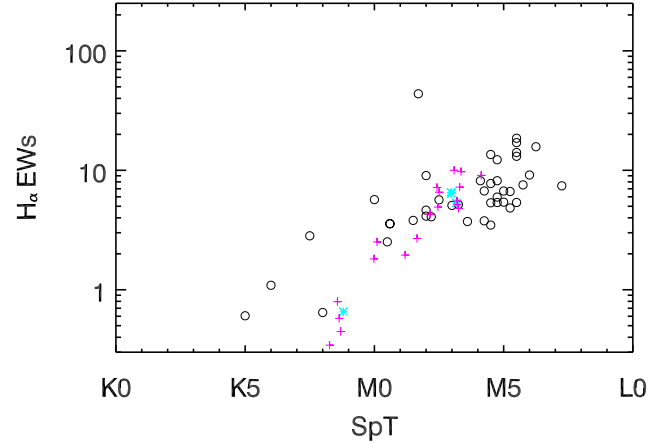}
\caption{The H${\alpha}$ Equivalent width of Tau-assocs members (cyan asterisks for u Tau assoc and purple crosses for e Tau assoc) compared with those of known Taurus PMS stars of no disk \cite[][shown as open black circles]{Esp2014}.}
\label{Halpha_EW}
\end{figure}

As the age of the Tau-assocs are estimated to be 50 Myr, stars under the TOn ($\sim 0.8 M_{\odot}$) should be post-T Tauri stars. To certificate the validity, spectra of the Tau-assocs observed by LAMOST is examined. In total 44 of them has been observed, their equivalent with of H${\alpha}$ line confirmed the transition from absorption to emission is roughly around the TOn. And the 18 stars with mass lower than the TOn mass do show obviously H${\alpha}$ emission feature. Compared their H${\alpha}$ equivalent width to that of previous known T Tauri stars with no circumstellar disk \citep{Esp2014}, it shows that the intensities are comparatively moderate to that of T Tauri stars (see figure \ref{Halpha_EW}), manifesting their youth properties.

\section{Summary and conclusions}\label{sum}



In this work, with {\em Gaia} DR2 astrometric data,
by searching over densities of nearby stars in the multi-phase space, we find 2 new young stellar associations, u Tau assoc and e Tau assoc, that haven't been noticed before. The two associations are quite close to
each other, but could be clearly separated.  Members of u Tau assoc are tightly concentrated in both $\alpha, \delta$ and $\mu_{\alpha}, \mu_{\delta}$ space, while the members of e Tau assoc, although also concentrated in the proper motions space, is more extended in the $\alpha$ and $\delta$ space within nearly 200 square degrees.
These two associations are of solar metallicity and young, with their best fitted isochrone ages are of about 50 Myr. From the fitting PMS model isochrones, the transition from Pre-main sequence to main sequence can be identified at $\sim$0.8~$M_{\odot}$, and  members lower than $\sim$0.8~$M_{\odot}$ are at the stage of post T~Tauri.




\section{acknowledgement}

We thank the anonymous referee for very helpful suggestions and comments. Quite a substantial data processing of this work are executed through the software of TOPCAT \citep{Tay2005}.
This work is supported by the National Natural Science Foundation of China (NSFC) with grant No. 11835057. Guoshoujing Telescope (the Large Sky Area Multi-Object Fiber Spectroscopic Telescope LAMOST) is a National Major Scientific Project built by the Chinese Academy of Sciences. Funding for the project has been provided by the National Development and Reform Commission. LAMOST is operated and managed by the National Astronomical Observatories, Chinese Academy of Sciences. This work has made use of data from the European Space Agency (ESA) mission {\it Gaia} (\url{https://www.cosmos.esa.int/gaia}), processed by the {\it Gaia} Data Processing and Analysis Consortium (DPAC, \url{https://www.cosmos.esa.int/web/gaia/dpac/consortium}). Funding for the DPAC has been provided by national institutions, in particular the institutions participating in the {\it Gaia} Multilateral Agreement.

\appendix
\section{Reliability test of ROCKSTAR}

\begin{figure}
\centering
\includegraphics[height=7cm]{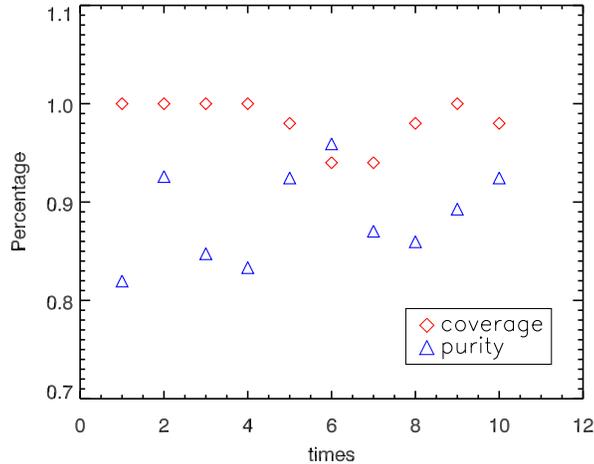}
\caption{The result of searching for the created association. 
Coverage means how many of the 50 group stars are located in the result. 
Purity is the group members left to that of total number of stars left in the process.}
\label{test}
\end{figure}

The purpose of ROCKSTAR we employed in this work is to refine the memberships 
and eliminated contaminations of the associations rather than discover associations, 
since we knew the two associations are clearly visible in the proper motion space. 
In considering that the ROCKSTAR code is not originally designed to handle associations,
therefore, a test about the effectiveness of ROCKSTAR is necessary. 
Given that young associations will be questioned about the completeness and contamination, 
a man-made artificial testing association will be much better. 
Thus, we randomly created a testing group of 50 stars of the same 
general properties of Tau-assocs (u Tau assoc and e Tau assoc) 
and a larger region of field stars surround them, 
also with likely properties as those field stars around Tau-assocs.
With same selecting criteria as searching for them and also run ROCKSTAR for 3 times.
We repeat this procedure 10 times, and showed the result in the figure \ref{test}.

The result show that in 10 times of testing, ROCKSTAR can find out $\sim 95\%$ of the group members 
at a purity level of $\sim 90\%$, proved that the code is an effective way in refining memberships of Tau-assocs kind. 
By this We mean that the ROCKSTAR code is an effective way in searching for members of Tau-assocs-like, but not for all the associations.

\clearpage
\begin{table}
\centering
\caption{Members of u Tau assoc.}
\label{LFL1}
\begin{tabular}{c|l|c|c|c|c|c|c|c}
\hline\hline\\
 GAIA ID &   other names &      RAJ2000     &           DEJ2000     &      pmRA    &       pmDEC     & Distance  &\ $\rm R_{V}$& Sp-Type\\
 &&deg & deg& mas/yr & mas/yr & pc &km/s & \\
\hline
   3275316947256508160 &                      &   51.736 &    4.043 &   21.629 &  -12.003 &  169.957 &         &            \\ 
   3274385076792329600 &                      &   54.059 &    3.927 &   23.147 &  -12.892 &  190.228 &         &            \\ 
   3274390707494059008 & TYC  67-1230-1      &   54.099 &    3.927 &   23.626 &  -13.737 &  171.170 &   17.960 &            \\ 
   3275975863959274752 &                      &   54.276 &    5.830 &   23.058 &  -14.917 &  185.967 &         &            \\ 
   3275991188402579584 &                      &   54.725 &    5.894 &   22.249 &  -13.610 &  180.059 &         &            \\ 
   3275104191756729728 &                      &   54.853 &    4.913 &   22.005 &  -13.117 &  192.334 &         &            \\ 
   3274743551942762624 & HD  22903            &   55.202 &    4.317 &   21.904 &  -12.699 &  179.276 &         & A1V        \\ 
   3274716064152131584 &                      &   55.248 &    3.977 &   23.079 &  -13.807 &  173.449 &         &            \\ 
   3274845153689271552 &                      &   55.338 &    4.515 &   22.543 &  -13.525 &  175.622 &         &            \\ 
   3274725994116512128 &                      &   55.358 &    4.169 &   22.856 &  -13.424 &  172.744 &         &            \\ 
   3276764076357778944 &                      &   55.542 &    6.225 &   21.869 &  -14.103 &  178.589 &         &            \\ 
   3271397600621973120 &                      &   55.752 &    3.056 &   21.255 &  -12.536 &  185.012 &         &            \\ 
   3277675674577524736 &                      &   55.892 &    7.032 &   21.395 &  -14.778 &  182.235 &         &            \\ 
   3271821286258516608 & HD  23248            &   55.916 &    4.213 &   21.191 &  -12.935 &  181.532 &         & A5II/II    \\ 
   3276862310851856512 & TYC  71-674-1       &   55.954 &    6.369 &   21.940 &  -14.034 &  180.029 &   16.700 &            \\ 
   3276494111894881536 &                      &   56.166 &    5.625 &   21.650 &  -13.876 &  176.280 &         &            \\ 
   3271752777237464960 & V* V1273 Tau         &   56.222 &    3.992 &   22.694 &  -13.444 &  173.999 &   12.480 & K2         \\ 
   3276798401738487808 &                      &   56.288 &    6.201 &   21.826 &  -14.176 &  176.509 &         &            \\ 
   3276386050517837696 &                      &   56.402 &    5.186 &   20.473 &  -14.314 &  187.407 &         &            \\ 
   3276605295710700032 & * u Tau          &   56.419 &    6.050 &   21.878 &  -13.646 &  187.893 &         & B3V        \\ 
   3276604544094119424 &                      &   56.421 &    6.043 &   21.272 &  -14.111 &  182.324 &         &            \\ 
   3276604544093968896 &                      &   56.428 &    6.051 &   21.108 &  -13.801 &  179.229 &         &            \\ 
   3276604922051089664 &                      &   56.435 &    6.058 &   19.950 &  -13.189 &  178.949 &   15.820 & G9         \\ 
   3276490121867896448 &                      &   56.441 &    5.734 &   22.133 &  -12.300 &  196.321 &   17.220 & G8         \\ 
   3276586333432639744 &                      &   56.523 &    5.885 &   21.409 &  -13.928 &  178.837 &         &            \\ 
   3276584478006772224 &                      &   56.531 &    5.803 &   21.416 &  -14.028 &  173.695 &         &            \\ 
   3276527406481433600 &                      &   56.745 &    5.569 &   21.463 &  -13.828 &  185.585 &         &            \\ 
   3276620315213926400 &                      &   56.878 &    6.125 &   21.276 &  -13.085 &  197.202 &         &            \\ 
   3276641446452946304 &                      &   56.959 &    6.297 &   20.959 &  -13.265 &  183.644 &         &            \\ 
   3276629386183056768 & TYC  71-542-1       &   56.987 &    6.269 &   21.191 &  -13.396 &  184.213 &   16.630 & G2         \\ 
   3276624227927031296 &                      &   57.131 &    6.178 &   20.870 &  -13.151 &  187.155 &         &            \\ 
   3270622066967091840 &                      &   57.343 &    2.706 &   23.208 &  -13.737 &  167.057 &   16.930 & K5         \\ 
   3276932847096564736 &                      &   57.601 &    6.041 &   20.242 &  -15.262 &  177.833 &         &            \\ 
   3273682832461023232 & BD+05  560  &   58.771 &    5.680 &   20.421 &  -13.529 &  200.909 &         & A5         \\ 
   3274101952549519232 &                      &   58.980 &    6.854 &   19.703 &  -12.866 &  194.851 &         &            \\ 
\hline\hline
\end{tabular}
\end{table}
\clearpage


\startlongtable
\begin{deluxetable*}{{c|l|c|c|c|c|c|c|c}}
\renewcommand{\tabcolsep}{0.05cm}
\tablecaption{Members of e Tau assoc\label{LFL2}}
\tablehead{
\colhead{GAIA ID} & \colhead{other names} & \colhead{RAJ2000} & 
\colhead{DEJ2000} & \colhead{pmRA} & 
\colhead{pmDEC} & \colhead{Distance} & 
\colhead{$\rm R_{V}$} & \colhead{Sp-Type}  \\ 
\colhead{} &\colhead{} & \colhead{(deg)} & \colhead{(deg)} & \colhead{(mas/yr)} & 
\colhead{(mas/yr)} & \colhead{(pc)} & \colhead{(km/s)} &\colhead{}} 
\startdata
      9509336766831744 &                      &   50.505 &    6.508 &   26.382 &  -22.743 &  133.808 &          &            \\ 
     10942515813999488 & HD  21194            &   51.362 &    8.427 &   26.855 &  -25.050 &  136.770 &   19.460 & F5         \\ 
      9993671639120256 & BD+06   533          &   51.991 &    7.255 &   27.982 &  -25.361 &  134.707 &          & F8         \\ 
     41872327659525504 &                      &   52.540 &   14.325 &   26.191 &  -24.738 &  148.766 &          &            \\ 
     42352367565698304 &                      &   53.199 &   15.473 &   24.905 &  -24.567 &  150.619 &          &            \\ 
     13078626388631424 & V* V1267 Tau         &   53.298 &   10.599 &   26.370 &  -25.552 &  136.872 &   15.080 & K3         \\ 
     11397988505713536 & HD  22073            &   53.444 &    8.291 &   27.040 &  -23.569 &  146.711 &          & A5         \\ 
     42440500294245120 &                      &   53.754 &   15.662 &   22.055 &  -23.234 &  156.706 &          &            \\ 
     11472617857575552 &                      &   53.869 &    8.199 &   27.438 &  -23.753 &  142.724 &          &            \\ 
     40695158727074048 &                      &   53.895 &   12.884 &   24.773 &  -24.208 &  163.725 &          &            \\ 
     40705848902910464 &                      &   54.027 &   13.128 &   27.849 &  -25.848 &  144.380 &          &            \\ 
     57153099745350912 &                      &   54.208 &   19.135 &   24.200 &  -24.869 &  150.991 &          &            \\ 
     40726224227818240 &                      &   54.306 &   13.125 &   25.209 &  -24.951 &  156.096 &          &            \\ 
     11522435182853120 &                      &   54.325 &    8.793 &   27.684 &  -24.865 &  142.689 &          &            \\ 
   3277897741565006976 &                      &   54.382 &    7.551 &   25.466 &  -24.476 &  145.419 &          &            \\ 
     44258027374647808 &                      &   54.396 &   17.088 &   23.681 &  -23.716 &  158.050 &   12.740 & K7         \\ 
     12343637226131584 & TYC  660-709-1       &   54.576 &   10.338 &   26.748 &  -24.923 &  146.051 &   14.560 & G9         \\ 
     12355628774864128 &                      &   54.704 &   10.363 &   25.724 &  -23.969 &  147.480 &          &            \\ 
     44034311118104320 &                      &   54.718 &   16.595 &   22.937 &  -25.649 &  158.159 &          &            \\ 
     11959323551830912 &                      &   54.793 &    9.466 &   24.434 &  -24.441 &  144.029 &   17.040 & K3         \\ 
     41651772500169984 & TYC 1235-156-1       &   54.915 &   15.499 &   25.957 &  -26.861 &  155.433 &   15.870 &            \\ 
     38088873789758720 & BD+12   500          &   55.042 &   13.199 &   25.527 &  -24.403 &  150.120 &   14.170 & F8         \\ 
     11888473771374848 &                      &   55.143 &    9.114 &   23.332 &  -24.108 &  147.665 &          &            \\ 
     37195348793250048 &                      &   55.154 &   11.293 &   27.113 &  -25.476 &  142.693 &          &            \\ 
     38076641722829440 & TYC  663-362-1       &   55.241 &   13.151 &   24.656 &  -25.443 &  147.577 &   13.360 &            \\ 
   3275164390018316288 &                      &   55.273 &    5.454 &   26.928 &  -23.628 &  141.578 &          &            \\ 
     11985196435903488 &                      &   55.283 &    9.285 &   24.988 &  -23.192 &  149.482 &          &            \\ 
     36420502331404160 & TYC  660-135-1       &   55.442 &   10.908 &   26.399 &  -25.756 &  140.121 &   14.500 &            \\ 
     44057778819482496 &                      &   55.515 &   16.528 &   24.365 &  -24.625 &  150.486 &          &            \\ 
     36103396308075776 &                      &   55.966 &   10.088 &   22.041 &  -24.349 &  150.297 &          &            \\ 
     36530487856056704 &                      &   56.072 &   11.303 &   26.253 &  -24.827 &  136.375 &          &            \\ 
   3277686910210391296 & 2MASS J03442859+0716 &   56.119 &    7.270 &   25.801 &  -22.564 &  150.709 &          & M4         \\ 
     36537737760831744 &                      &   56.126 &   11.501 &   25.259 &  -24.176 &  149.959 &          &            \\ 
     37844713488859264 &                      &   56.147 &   12.959 &   24.614 &  -24.007 &  154.850 &          &            \\ 
     36034023996411392 &                      &   56.179 &    9.738 &   22.636 &  -24.009 &  150.274 &          &            \\ 
   3278197770802258944 & HD  23376            &   56.246 &    8.320 &   26.612 &  -24.306 &  144.644 &   16.490 & G5         \\ 
   3278197766505583232 & TYC  658-1007-2      &   56.246 &    8.321 &   26.577 &  -24.198 &  142.418 &   20.640 &            \\ 
   3278487148518773248 &                      &   56.306 &    8.616 &   26.363 &  -23.826 &  155.014 &          &            \\ 
   3278300987456845440 & TYC  658-828-1       &   56.467 &    8.541 &   27.783 &  -24.644 &  131.926 &   15.590 &            \\ 
   3278489313182286720 &                      &   56.501 &    8.609 &   27.791 &  -24.834 &  135.925 &          &            \\ 
     50970717660507008 &                      &   56.690 &   19.189 &   25.563 &  -23.138 &  154.909 &          &            \\ 
   3277287615693093376 &                      &   56.717 &    7.078 &   27.049 &  -22.842 &  150.719 &          &            \\ 
   3277313144978675840 &                      &   56.726 &    7.344 &   25.167 &  -22.944 &  142.686 &          &            \\ 
   3302676747927557504 &                      &   56.741 &    9.945 &   27.262 &  -27.025 &  137.566 &          &            \\ 
   3276420135378285056 &                      &   56.798 &    5.440 &   22.837 &  -19.710 &  163.312 &          &            \\ 
     42956995881088256 &                      &   56.847 &   15.557 &   21.727 &  -24.897 &  157.735 &          &            \\ 
     43660752042391680 &                      &   56.849 &   16.808 &   24.806 &  -24.601 &  154.295 &          &            \\ 
   3278204402232390528 & BD+07   543          &   56.880 &    7.957 &   25.369 &  -24.690 &  155.451 &   14.640 & F8         \\ 
     36203417507399808 &                      &   56.966 &   10.724 &   27.642 &  -25.464 &  158.885 &          &            \\ 
     36901298152686080 & TYC  661-560-1       &   56.974 &   11.816 &   21.230 &  -25.115 &  154.141 &   15.450 &            \\ 
     43059486684334208 &                      &   57.043 &   16.145 &   23.268 &  -25.243 &  159.908 &          &            \\ 
     36290072767519232 & * e Tau              &   57.068 &   11.143 &   25.269 &  -23.695 &  128.770 &          & B3V        \\ 
     36290107127257344 & TYC  661-1404-1      &   57.070 &   11.145 &   25.616 &  -24.972 &  137.914 &   19.560 & F3Vn       \\ 
   3278259858850059264 & TYC  658-922-1       &   57.131 &    8.527 &   25.327 &  -22.738 &  151.878 &    8.520 & G7         \\ 
     36941189808895872 &                      &   57.184 &   12.220 &   23.753 &  -23.215 &  159.044 &   11.880 & K7         \\ 
   3276628183593979904 &                      &   57.218 &    6.342 &   23.643 &  -20.334 &  155.848 &          &            \\ 
   3277330153049126400 &                      &   57.228 &    7.465 &   24.847 &  -22.347 &  150.250 &          &            \\ 
   3302817966452511616 &                      &   57.425 &   10.591 &   24.606 &  -24.003 &  149.917 &          &            \\ 
   3302396166303947904 & HD  23990            &   57.444 &    9.407 &   25.165 &  -24.660 &  147.495 &          & B9         \\ 
     39846683645349376 &                      &   57.460 &   14.682 &   22.024 &  -24.158 &  154.757 &          &            \\ 
     36924353537157632 &                      &   57.522 &   12.071 &   24.224 &  -23.552 &  153.293 &          &            \\ 
     36701702432783616 &                      &   57.572 &   11.496 &   24.337 &  -24.516 &  150.262 &          &            \\ 
     43458167025621504 & 2MASS J03502840+1631 &   57.618 &   16.521 &   24.242 &  -21.892 &  146.359 &    7.390 & G5IV       \\ 
     39513359823463680 &                      &   57.620 &   13.937 &   23.942 &  -24.043 &  149.477 &          &            \\ 
   3302299134402909056 &                      &   57.638 &    8.930 &   24.037 &  -22.827 &  154.236 &          &            \\ 
     36724620378249984 &                      &   57.683 &   11.809 &   22.542 &  -22.481 &  165.514 &          &            \\ 
     36595943156045824 & BD+10   496          &   57.711 &   11.002 &   24.137 &  -24.167 &  151.392 &   14.960 & F8         \\ 
   3302822811175649664 &                      &   57.724 &   10.702 &   26.015 &  -24.606 &  141.009 &          &            \\ 
     39641487287397632 &                      &   57.772 &   14.526 &   20.639 &  -23.587 &  154.918 &          &            \\ 
   3301516179044339840 &                      &   57.786 &    8.489 &   25.935 &  -24.427 &  137.591 &          &            \\ 
     37136524921755904 &                      &   57.793 &   13.046 &   23.101 &  -23.138 &  162.124 &          &            \\ 
     37136834159399808 & V* V766 Tau          &   57.816 &   13.046 &   23.769 &  -23.228 &  160.771 &          & B9pSi      \\ 
     39841357885932288 & TYC  664-136-1       &   57.915 &   14.797 &   23.833 &  -23.721 &  159.686 &   28.720 &            \\ 
   3277156567650183680 &                      &   58.149 &    7.156 &   21.228 &  -22.856 &  158.009 &          &            \\ 
   3270343546928113664 & HD  24456            &   58.376 &    2.119 &   26.933 &  -20.785 &  138.680 &          & B9V        \\ 
   3273850404904742912 & TYC   72-816-1       &   58.449 &    5.707 &   28.184 &  -25.182 &  132.266 &   27.560 &            \\ 
   3270377941026192768 &                      &   58.506 &    2.484 &   26.104 &  -25.683 &  124.607 &          &            \\ 
   3273648919399332608 & TYC   72-620-1       &   58.513 &    5.431 &   24.625 &  -20.253 &  142.703 &   22.210 &            \\ 
   3273169120012500736 &                      &   58.707 &    4.624 &   25.847 &  -22.798 &  141.751 &          &            \\ 
   3301633517550972032 &                      &   58.811 &    8.800 &   20.422 &  -24.161 &  154.244 &          &            \\ 
   3302850402044815104 &                      &   58.836 &    9.921 &   23.750 &  -23.323 &  151.633 &          &            \\ 
   3301687599779043584 &                      &   58.933 &    9.301 &   25.201 &  -24.605 &  149.626 &          &            \\ 
   3303308245556503296 & HD 286374            &   59.080 &   11.420 &   24.048 &  -24.124 &  152.306 &   13.600 & F5         \\ 
   3303061851874905088 & HD 286380            &   59.086 &   10.797 &   24.658 &  -25.252 &  147.681 &   14.650 & G5         \\ 
   3302885135443350144 &                      &   59.206 &   10.174 &   23.899 &  -24.142 &  150.217 &          &            \\ 
     38398936068862464 & TYC  665-150-1       &   59.339 &   12.971 &   28.046 &  -24.709 &  152.421 &          & G0         \\ 
   3303319927869595264 &                      &   59.412 &   11.709 &   23.298 &  -23.208 &  154.757 &          &            \\ 
   3273771824183136256 &                      &   59.428 &    5.856 &   28.109 &  -25.250 &  126.075 &          &            \\ 
   3302063667115147008 & 2MASS J03581272+0932 &   59.553 &    9.540 &   24.321 &  -24.493 &  146.782 &   16.120 & K3         \\ 
   3273802232551662336 &                      &   59.748 &    6.092 &   24.363 &  -22.545 &  150.096 &          &            \\ 
   3272119941104628352 &                      &   59.750 &    3.837 &   23.782 &  -21.905 &  156.004 &          &            \\ 
   3259900660364779392 &                      &   59.767 &    2.881 &   21.447 &  -21.952 &  142.016 &          &            \\ 
   3301312838112630400 &                      &   59.814 &    8.289 &   25.603 &  -25.603 &  149.855 &          &            \\ 
   3304906145189468416 & TYC  662-217-1       &   59.926 &   12.169 &   24.066 &  -25.242 &  148.495 &   15.290 &            \\ 
   3304619413175027968 &                      &   60.004 &   11.611 &   24.188 &  -25.297 &  146.915 &          &            \\ 
   3301396126118384768 &                      &   60.013 &    8.653 &   23.291 &  -22.675 &  158.536 &          &            \\ 
   3259830325980399744 &                      &   60.132 &    2.593 &   23.916 &  -20.774 &  152.736 &          &            \\ 
   3302018999455336192 &                      &   60.158 &    9.367 &   22.724 &  -22.632 &  157.019 &          &            \\ 
   3301329949262394624 &                      &   60.292 &    8.406 &   23.114 &  -22.334 &  152.465 &          &            \\ 
   3301831773241303552 & BD+08   616          &   60.339 &    9.334 &   25.238 &  -26.184 &  155.343 &   42.580 & F8         \\ 
   3272433615452990848 &                      &   60.347 &    3.709 &   24.897 &  -22.773 &  140.743 &          &            \\ 
   3298319348986238464 &                      &   60.553 &    8.294 &   23.378 &  -22.681 &  150.586 &          &            \\ 
   3301945366536236416 &                      &   60.698 &    9.776 &   20.112 &  -25.812 &  146.839 &          &            \\ 
   3297800516936921344 &                      &   60.738 &    6.383 &   27.282 &  -26.218 &  135.379 &  -35.940 &            \\ 
   3297045667844723712 &                      &   60.893 &    6.298 &   23.739 &  -22.357 &  142.644 &          &            \\ 
   3298606905637432576 &                      &   61.165 &    7.866 &   26.820 &  -27.881 &  158.535 &          &            \\ 
   3301900595797205632 &                      &   61.206 &    9.585 &   22.404 &  -24.313 &  150.938 &          &            \\ 
   3301974331795653888 &                      &   61.211 &    9.936 &   22.900 &  -23.903 &  150.802 &          &            \\ 
   3297959396367266560 &                      &   61.304 &    7.172 &   22.500 &  -21.909 &  157.938 &          &            \\ 
   3297032886022084864 &                      &   61.344 &    6.259 &   23.325 &  -22.646 &  153.044 &          &            \\ 
   3296973134437145984 &                      &   61.436 &    6.014 &   27.429 &  -25.740 &  161.976 &          &            \\ 
   3298826700586330240 &                      &   61.667 &    8.941 &   23.444 &  -24.541 &  138.867 &          &            \\ 
   3297062675915933696 &                      &   61.753 &    6.119 &   27.861 &  -26.029 &  119.267 &          &            \\ 
   3298371507070141056 & TYC  666-80-1        &   62.448 &    7.801 &   21.622 &  -22.284 &  157.801 &   19.060 &            \\ 
   3297619097518627968 & HD  26323            &   62.529 &    7.698 &   22.376 &  -20.975 &  161.069 &          & B9         \\ 
   3297666204719373440 &                      &   63.798 &    7.764 &   20.502 &  -21.183 &  163.552 &          &            \\ 
   3299306676067811200 & * mu. Tau            &   63.884 &    8.892 &   20.881 &  -22.789 &  149.992 &          & B3IV       \\ 
   3300180959610801536 &                      &   63.944 &    9.357 &   19.986 &  -21.413 &  163.528 &          &            \\ 
   3285243784909511168 &                      &   64.650 &    6.244 &   22.561 &  -21.392 &  159.112 &          &            \\ 
\enddata
\end{deluxetable*}
\clearpage


\begin{thebibliography}{}

\bibitem[Bailer-Jones et al.(2018)]{Bai2018} Bailer-Jones, C.~A.~L., Rybizki, J., Fouesneau, M., Mantelet, G., \& Andrae, R.\ 2018, \aj, 156, 58 

\bibitem[Behroozi et al.(2013)]{Beh2013} Behroozi, P.~S., Wechsler, R.~H., \& Wu, H.-Y.\ 2013, \apj, 762, 109
\bibitem[Bell et al.(2014)]{Bel2014} Bell, C.~P.~M., Rees, J.~M., Naylor, T., et al.\ 2014, \mnras, 445, 3496 

\bibitem[Bressan et al.(2012)]{Bre2012} Bressan, A., Marigo, P., Girardi, L., et al.\ 2012, \mnras, 427, 127 
\bibitem[Brice{\~n}o et al.(1998)]{Bri1998} Brice{\~n}o, C., Hartmann, L., Stauffer, J., \& Mart{\'{\i}}n, E.\ 1998, \aj, 115, 2074
\bibitem[Brice{\~n}o et al.(1999)]{Bri1999} Brice{\~n}o, C., Calvet, N., Kenyon, S., \& Hartmann, L.\ 1999, \aj, 118, 1354 
\bibitem[Brice{\~n}o et al.(2002)]{Bri2002} Brice{\~n}o, C., Luhman, K.~L., Hartmann, L., Stauffer, J.~R., \& Kirkpatrick, J.~D.\ 2002, \apj, 580, 317 

\bibitem[Cardelli et al.(1989)]{Car1989} Cardelli, J.~A., Clayton, G.~C., \& Mathis, J.~S.\ 1989, \apj, 345, 245

\bibitem[Chauvin et al.(2015)]{Chauvin2015} Chauvin, G., Vigan, A., Bonnefoy, M., et al.\ 2015, \aap, 573, A127.

\bibitem[Chambers et al.(2016)]{Cha2016} Chambers, K.~C., Magnier, E.~A., Metcalfe, N., et al.\ 2016, arXiv:1612.05560 
\bibitem[Cignoni et al.(2010)]{Cig2010} Cignoni, M., Tosi, M., Sabbi, E., et al.\ 2010, \apjl, 712, L63 

\bibitem[Cui et al.(2012)]{Cui2012} Cui, X.-Q., et al. 2012, Research in Astronomy and Astrophysics,12, 1197
\bibitem[de Bruijne(1999)]{de1999} de Bruijne, J.~H.~J.\ 1999, \mnras, 306, 381

\bibitem[Dotter et al.(2008)]{Dor2008} Dotter, A., Chaboyer, B., Jevremovi{\'c}, D., et al.\ 2008, \apjs, 178, 89

\bibitem[Esplin et al.(2014)]{Esp2014} Esplin, T.~L., Luhman, K.~L., \& Mamajek, E.~E.\ 2014, \apj, 784, 126
\bibitem[Esplin \& Luhman(2017)]{Esp2017a} Esplin, T.~L., \& Luhman, K.~L.\ 2017, \aj, 154, 134 
\bibitem[Esplin et al.(2017)]{Esp2017b} Esplin, T.~L., Luhman, K.~L., Faherty, J.~K., Mamajek, E.~E., \& Bochanski, J.~J.\ 2017, \aj, 154, 46 

\bibitem[Faherty et al.(2018)]{Fah2018} Faherty, J.~K., Bochanski, J.~J., Gagn{\'e}, J., et al.\ 2018, \apj, 863, 91 

\bibitem[Fang et al.(2017)]{Fan2017} Fang, M., Kim, J.~S., Pascucci, I., et al.\ 2017, \aj, 153, 188 
\bibitem[Feiden et al.(2015)]{Fei2015} Feiden, G.~A., Jones, J., \& Chaboyer, B.\ 2015, 18th Cambridge Workshop on Cool Stars, Stellar Systems, and the Sun, 18, 171 

\bibitem[F{\"u}rnkranz et al.(2019)]{Fur2019} F{\"u}rnkranz, V., Meingast, S., \& Alves, J.\ 2019, \aap, 624, L11

\bibitem[Gagn{\'e} et al.(2018)]{Gag2018} Gagn{\'e}, J., Mamajek, E.~E., Malo, L., et al.\ 2018, \apj, 856, 23 
\bibitem[Galli et al.(2012)]{Gal2012} Galli, P.~A.~B., Teixeira, R., Ducourant, C., Bertout, C., \& Benevides-Soares, P.\ 2012, \aap, 538, A23 

\bibitem[Galli et al.(2018)]{Gal2018} Galli, P.~A.~B., Loinard, L., Ortiz-L{\'e}on, G.~N., et al.\ 2018, \apj, 859, 33 

\bibitem[Gaia Collaboration et al.(2016)]{GAIA2016} Gaia Collaboration et al., 2016, A\&A, 595, A2
\bibitem[Gaia Collaboration et al.(2017)]{GAIA2017} Gaia Collaboration et al., 2017, A\&A, 601, A19
\bibitem[Gaia Collaboration et al.(2018)]{GAIA2018} Gaia Collaboration, Brown, A.~G.~A., Vallenari, A., et al.\ 2018, \aap, 616, A1 

\bibitem[Gottlieb and Upson(1969)]{Got1969} Gottlieb, D.~M., \& Upson, W.~L.\ 1969, \apj, 157, 611
\bibitem[Guieu et al.(2006)]{Gui2006} Guieu, S., Dougados, C., Monin, J.-L., Magnier, E., \& Mart{\'{\i}}n, E.~L.\ 2006, \aap, 446, 485
\bibitem[Herbig(1965)]{Her1965} Herbig, G.~H.\ 1965, \apj, 141, 588 
\bibitem[Herbig(1978)]{Her1978} Herbig, G.~H.\ 1978, Problems of Physics and Evolution of the Universe, 171 
\bibitem[Jensen(2001)]{Jen2001} Jensen, E.~L.~N.\ 2001, Young Stars Near Earth: Progress and Prospects, 244, 3 

\bibitem[Joncour et al.(2017)]{Jon2017} Joncour, I., Duch{\^e}ne, G., \& Moraux, E.\ 2017, \aap, 599, A14
\bibitem[Jones(1971)]{Jon1971} Jones, D.~H.~P.\ 1971, \mnras, 152, 231 
\bibitem[Kraus et al.(2017)]{Kra2017} Kraus, A.~L., Herczeg, G.~J., Rizzuto, A.~C., et al.\ 2017, \apj, 838, 150
\bibitem[Kroupa(2001)]{Kro2001} Kroupa, P.\ 2001, \mnras, 322, 231 
\bibitem[Lim et al.(2016)]{Lim2016} Lim, B., Sung, H., Kim, J.~S., et al.\ 2016, \apj, 831, 116 

\bibitem[Luhman(2004)]{Luh2004} Luhman, K.~L.\ 2004, \apj, 617, 1216.
\bibitem[Luhman(2007)]{Luh2007} Luhman, K.~L.\ 2007, \apjs, 173, 104.
\bibitem[Luhman et al.(2009)]{Luh2009} Luhman, K.~L., Mamajek, E.~E., Allen, P.~R., \& Cruz, K.~L.\ 2009, \apj, 703, 399
\bibitem[Luhman et al.(2017)]{Luh2017} Luhman, K.~L., Mamajek, E.~E., Shukla, S.~J., \& Loutrel, N.~P.\ 2017, \aj, 153, 46
\bibitem[Luhman(2018)]{Luh2018} Luhman, K.~L.\ 2018, \aj, 156, 271
\bibitem[Luo et al.(2012)]{Luo2012} Luo, A.-L., Zhang, H.-T., Zhao, Y.-H., et al.\ 2012, Research in Astronomy and Astrophysics, 12, 1243 

\bibitem[Milne and Aller(1980)]{Mil1980} Milne, D.~K., \& Aller, L.~H.\ 1980, \aj, 85, 17
\bibitem[Murphy \& Lawson(2015)]{Mur2015} Murphy, S.~J., \& Lawson, W.~A.\ 2015, \mnras, 447, 1267 

\bibitem[Oh et al.(2017)]{Oh2017} Oh, S., Price-Whelan, A.~M., Hogg, D.~W., Morton, T.~D., \& Spergel, D.~N.\ 2017, \aj, 153, 257 

\bibitem[Ortega et al.(2009)]{Ort2009} Ortega, V.~G., Jilinski, E., de la Reza, R., \& Bazzanella, B.\ 2009, \aj, 137, 3922 
\bibitem[R{\"o}ser et al.(2018)]{Ros2018} R{\"o}ser, S., Schilbach, E., Goldman, B., et al.\ 2018, \aap, 614, A81

\bibitem[Smart(1968)]{Sma1968} Smart, W.~M.\ 1968, London: Longmans, 1968,  

\bibitem[Song et al.(2003)]{Son2003} Song, I., Zuckerman, B., \& Bessell, M.~S.\ 2003, \apj, 599, 342

\bibitem[Stauffer(1980)]{Sta1980} Stauffer, J.~R.\ 1980, \aj, 85, 1341 

\bibitem[Su \& Cui(2004)]{Su2004} Su, D.-Q., \& Cui, X.-Q.\ 2004, \cjaa, 4, 1 

\bibitem[Taylor(2005)]{Tay2005} Taylor, M.~B.\ 2005, Astronomical Data Analysis Software and Systems XIV, 29

\bibitem[Tetzlaff et al.(2011)]{Tet2011} Tetzlaff, N., Neuh{\"a}user, R., \& Hohle, M.~M.\ 2011, \mnras, 410, 190 
\bibitem[Tognelli et al.(2012)]{Ton2012} Tognelli, E., Degl'Innocenti, S., \& Prada Moroni, P.~G.\ 2012, \aap, 548, A41

\bibitem[Tonry et al.(2012)]{Tonr2012} Tonry, J.~L., Stubbs, C.~W., Lykke, K.~R., et al.\ 2012, \apj, 750, 99 

\bibitem[Torres et al.(2008)]{Tor2008} Torres, G., Winn, J.~N., \& Holman, M.~J.\ 2008, \apj, 677, 1324 

\bibitem[Walter(1986)]{Wal1986} Walter, F.~M.\ 1986, \apj, 306, 573 

\bibitem[Wang et al.(2017)]{Wan2017} Wang, S., Jiang, B.~W., Zhao, H., et al.\ 2017, \apj, 848, 106
\bibitem[Wang and Chen(2019)]{Wan2019} Wang, S., \& Chen, X.\ 2019, \apj, 877, 116


\bibitem[Weymann \& Sears(1965)]{Wey1965} Weymann, R., \& Sears, R.~L.\ 1965, \apj, 142, 174 

\bibitem[Wichmann et al.(1998)]{Wic1998} Wichmann, R., Bastian, U., Krautter, J., Jankovics, I., \& Rucinski, S.~M.\ 1998, \mnras, 301, L39 

\end{thebibliography}
\end{document}